\documentclass[twocolumn]{webofc}

\usepackage[varg]{txfonts}
\usepackage{units}

\begin{document}
\title{Review of recent forward physics
results from the CMS experiment}

\author{\firstname{Ralf Ulrich} \inst{1}\fnsep\thanks{\email{ralf.m.ulrich@kit.edu}} for the CMS Collaboration}

\institute{Institute for Nuclear Physics, Karlsruhe Institute of Technology (KIT), 76131 Karlsruhe} 

\abstract{There is a rich program of forward physics measurements
  within the CMS Collaboration covering a wide range of topics. In many
  cases there is a connection to quantities and
  effects relevant for very high energy cosmic ray interaction. Some of the
  recent measurements in the fields of exclusive final states,
  low-p$_{\rm T}$ inclusive and diffractive cross sections, underlying event,
  multi parton interactions, double parton scattering, final state
  particle correlations and minimum bias results are briefly summarized here. }

\maketitle

\section{Introduction}

In this proceedings some of the recent results of the forward physics
activity of the CMS Collaboration are summarized. It is interesting
that these measurements are not just limited to small (forward) angles,
but in particular cover also low-$p_{\rm T}$, exclusive final states
and in particular minimum bias results.
The type of measurements discussed here cover up to 14 orders of
magnitude in cross section, ranging from very rare exclusive
W$^+$W$^-$ pair production in proton-proton collisions at
$\sqrt{s}=$8\,TeV with a cross section of about $\approx10\,$fb$^{-1}$
up to the inclusive cross section in proton-lead collisions at
$\sqrt{s_{\rm NN}}=$5.02\,TeV of more then 2\,b.

Some of the measurements described have a  
relation to very high energy cosmic ray interactions and can be of use to
further improve the understanding of the modelling of those.

\section{CMS and its forward instrumentation}

The CMS Collaboration benefits from a powerful general purpose particle
experiment, with good forward instrumentation using dedicated detectors.

At the heart of the CMS detector is a superconducting solenoid of
6\unit{m} internal diameter, providing a strong magnetic field of
3.8\unit{T}. The data used for this paper were taken in June 2015
during a period without magnetic field.  Within the CMS magnet volume
are an inner silicon pixel and strip tracker that measure charged
particles in the range $|\eta|<2.5$, a homogeneous lead tungstate
crystal electromagnetic calorimeter, and a brass and scintillator
hadron calorimeter.  The corresponding endcap detectors instrument the
pseudorapidity range up to $|\eta|\lesssim3$ with tracking and
calorimetry.  Forward Cherenkov calorimeters extend the coverage
beyond $|\eta|\gtrsim3$.  Muons are measured in gas-ionization
detectors embedded in the steel return yoke.

The hadron forward (HF) calorimeters cover the region $2.9<|\eta|<5.2$
and consist of 2$\times$432 readout towers, each containing a long and
a short quartz fiber embedded within a steel absorber running parallel
to the beam. The long fibers run the entire depth of the HF
calorimeter (165\unit{cm}, or approximately 10 interaction length),
while the short fibers start at a depth of 22\unit{cm} from the front
of the detector. The response of each tower is determined from the sum
of signal in the corresponding long and short fiber.

\begin{figure}[pb!]
\includegraphics[width=\columnwidth]{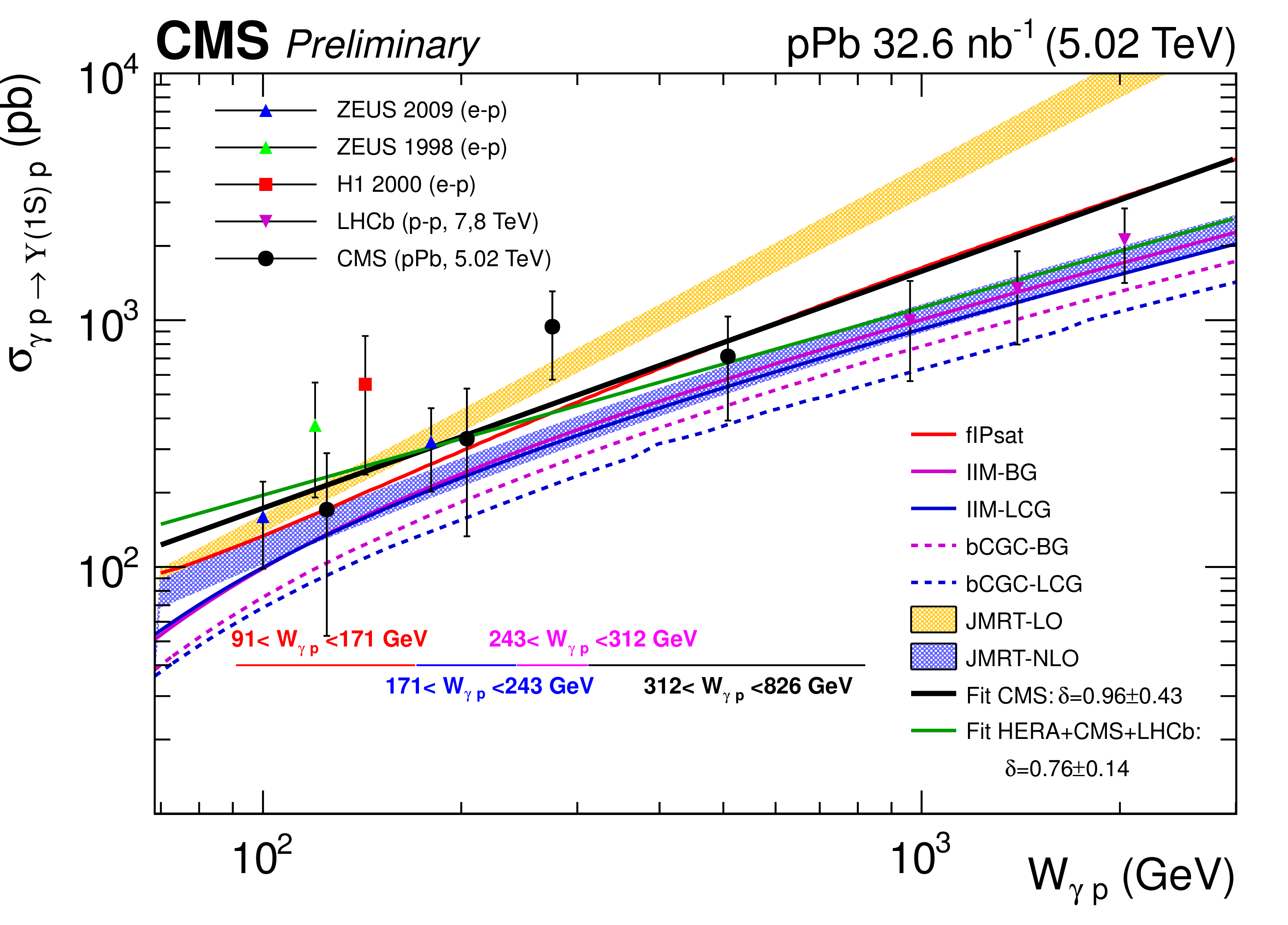}
\caption{Measurement of the photon-proton cross section as function of
  photon energy (from Ref.~\cite{Sirunyan:2018sav}). This result is
  part of the exclusive Y measurement by CMS in proton-lead collisions
  at $\sqrt{s_{\rm NN}}=5.02$\,TeV. }
\label{fig:exclY}
\end{figure}

The very forward angles on one side of CMS ($-6.6<\eta<-5.2$) are
covered by the CASTOR calorimeter.  It has 16 azimuthal towers, each
built from 14 longitudinal modules. The 2 front modules form the
electromagnetic section, and the 12 rear modules form the hadronic
section.  The calorimeter is made of stacks of tungsten and quartz
plates, read out by PMTs, in two half-cylindrical mechanical
structures, and is placed around the beam pipe at a distance of
$-14.4$\,m away from the nominal interaction point.  The overall
longitudinal depth of both CASTOR and HF corresponds to 10 hadronic
interaction lengths.

A more detailed description of the CMS detector can be found in
Ref.~\cite{Chatrchyan:2008aa}.

\section{Exclusive final states}

The exclusive production of a particular final state in a
hadron-hadron collision is only possible if the exchange particle is
colour neutral, thus, either a colour neutral hadronic state (i.e.\ a
pomeron), a photon, or a combination of both.

In particular the collisions where a pomeron is involved offer the
exciting opportunity to directly probe the low-x part of the gluon
distribution in the hadrons. The most prominent of such measurements
are the exclusive production of vector mesons. CMS has measured
exclusive Y production in proton-lead collisions at $\sqrt{s_{\rm
    NN}}$=5.02\,TeV. In this system the lead nucleus acts as the
source for the photons, while the proton for the pomerons. One result of
this measurement is the photon-proton cross section as a function of
the photon energy over a much wider range than previously possible,
see Fig.~\ref{fig:exclY}.

\begin{figure}[pt!]
\includegraphics[width=\columnwidth]{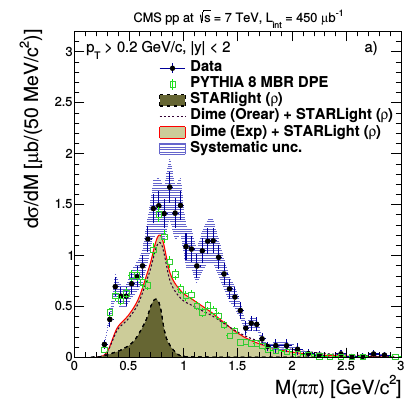}
\caption{The distribution of the invariant $\pi^+\pi^-$ mass
  distribution in (semi-) exclusive production, see
  Ref.~\cite{Khachatryan:2017xsi}, measured in proton-proton
  collisions at 7\,TeV. }
\label{fig:pipi}
\end{figure}

The (semi-)exclusive production of $\pi^+\pi^-$ in proton-proton
collisions at 7\,TeV is a much more complex environment. It contains a
continuum distribution, and resonances on top of that (see
Fig.~\ref{fig:pipi}). The main production mechanism is via
diffraction, but with a transition to the regime where one or both of
the protons dissociate and produce further particles. These data can
reveal information on the probability of proton dissociation in
diffractive interactions.

\begin{figure}[pt!]
\includegraphics[width=\columnwidth]{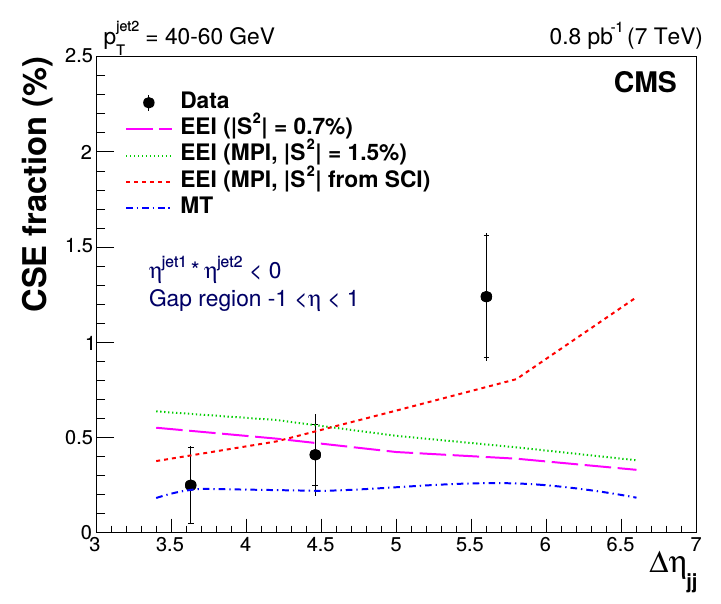}
\caption{The fraction of colour singlet exchange in events with two
  jets, see Ref.~\cite{Sirunyan:2017rdp}, as function of the
  jet-separation $\Delta\eta_{jj}$.}
\label{fig:rapgap}
\end{figure}

Furthermore, in inelastic collision where jets are produced it is
possible that the two jets are repelled from each other by a color
singlet exchange (CSE). The signature of this is studied in
jet-gap-jet measurements in proton-proton collisions at 7\,TeV, see
Fig.~\ref{fig:rapgap}. These data are useful to search for the
signature of BFKL effects.

\begin{figure}[pt!]
\includegraphics[width=\columnwidth]{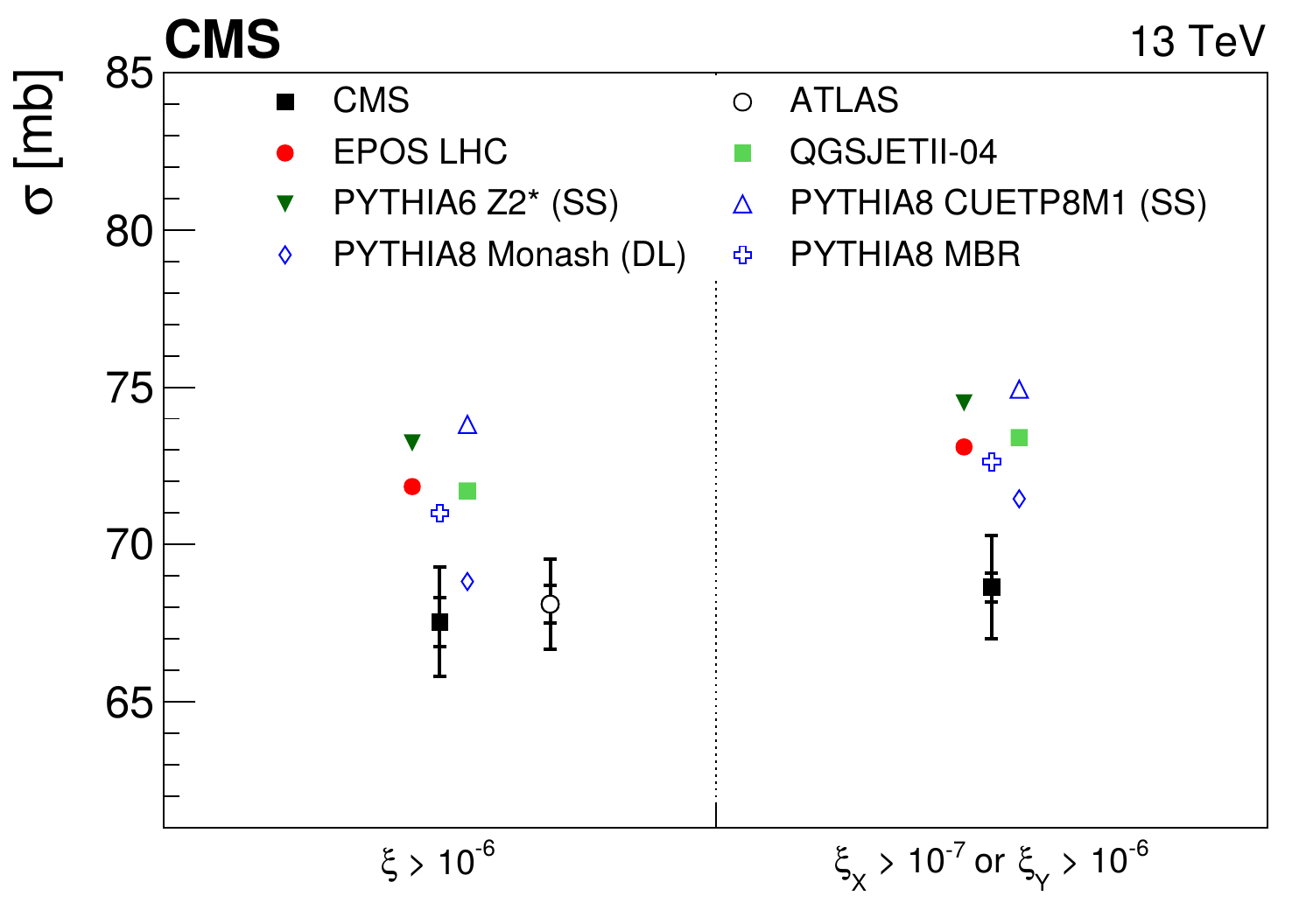}
\caption{The detector-level inelastic cross section in proton-proton
  collisions measured for $\xi>10^{-6}$ and for $\xi_x>10^{-7}$ or
  $\xi_y>10^{-6}$, see Ref.~\cite{Sirunyan:2018nqx}}
\label{fig:cx}
\end{figure}




\begin{figure*}[pbt!]
\includegraphics[width=\textwidth]{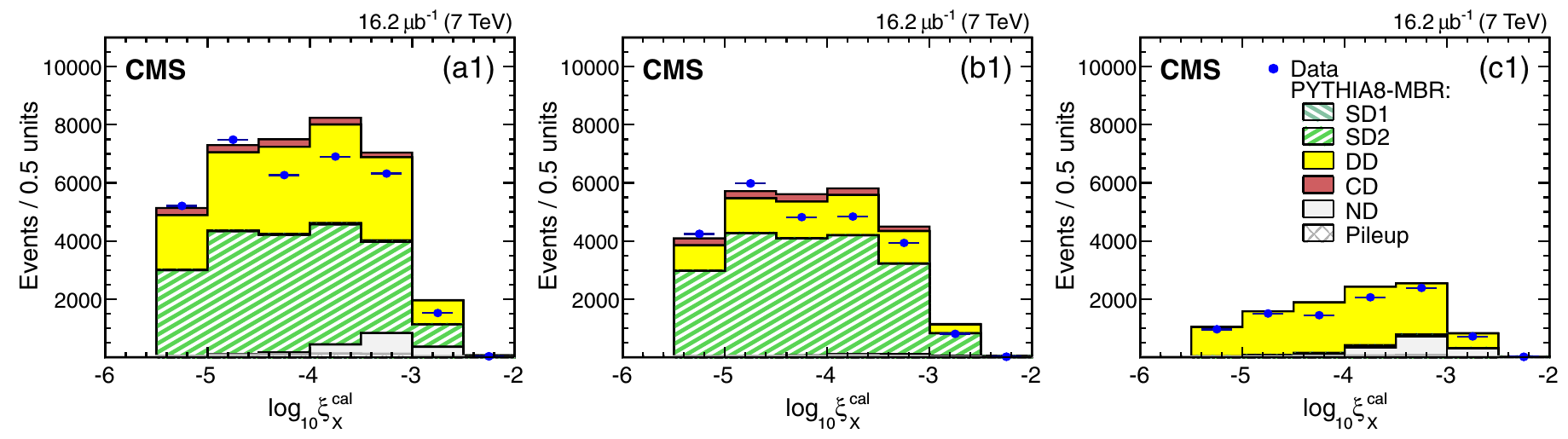}
  \caption{The shapes of the distributions of the number of events as
    function of the proton rapidity loss as determined by CMS
    calorimeters $\xi_x^{\rm cal}$. In (a1) all events with forward
    rapidity gap are shown, in (b1) all of those events with no signal
    in CASTOR, and in (c1) those events that were flagged by
    CASTOR. See Ref.~\cite{Khachatryan:2015gka}.}
\label{fig:SD}
\end{figure*}

Another type of collisions is where only photons are exchanged. At a
hadron collider this can provide an incredibly clean environment for
very rare processes. Namely, these are exclusive pair production where
e.g.\ W$^+$W$^-$ are a powerful probe for anomalous quartic gauge
coupling and physics beyond the standard
model~\cite{Khachatryan:2016mud}. If this can be combined with the
simultaneous observation of the two outgoing proton projectiles the
kinematics of the protons can be correlated with the central exclusive
system opening up a window for searches for new physics. The CMS-TOTEM
Proton Precision Spectrometer (CT-PPS) is build for this
purpose~\cite{Cms:2018het}.

\section{Inclusive cross sections, and diffraction}

One of the most important parameter concerning the development of
extensive air shower cascades is the inelastic cross section for
particle production. The inelastic cross section in proton-proton
collision has been measured at 13\,TeV~\cite{Sirunyan:2018nqx}. This
analysis is published in two acceptance regions, where the difference
is in including the CASTOR very forward calorimeter, see
Fig.~\ref{fig:cx}. It is very interesting to see that both data points
are considerable smaller than all model predictions. The CMS
measurement at $\xi>10^{-6}$ is consistent with the corresponding
measurement by ATLAS. A smaller inelastic cross section will lead to a
deeper penetration of air shower cascades.

In the determination of the inelastic cross section, the most
difficult problem is to understand the impact of low-mass
diffraction. The amount of cross section lost due to detector
acceptance is a relatively poorly constraint quantity. In
Fig.~\ref{fig:SD} the measured proton momentum loss, $\xi$, distributions in proton-proton
collisions at 7\,TeV are shown with different event selections. It can
be seen that the separation of single- and double-diffraction is
extremely difficult with typical LHC detectors. In CMS, only the
inclusion of the CASTOR calorimeter helps to get a better handle on
it.

\section{Measurements of inclusive distributions}

\begin{figure}
\includegraphics[width=\columnwidth]{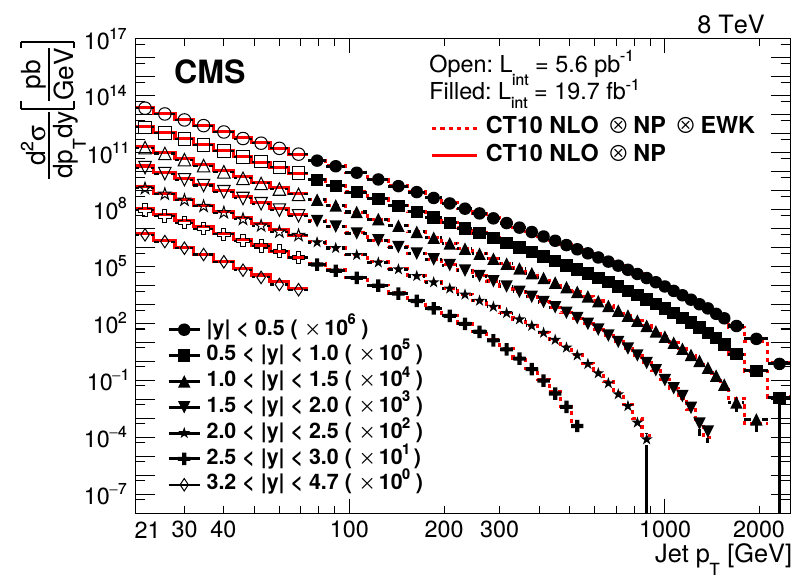}
\caption{Differential cross sections of inclusive jet production as
  function of jet $p_{\rm T}$, see
  Ref.~\cite{Khachatryan:2016mlc}. The filled markers are measured by
  all CMS Calorimeters with $\eta<3$, while the open markers by the HF
  calorimeters.}
\label{fig:jet}
\end{figure}

The measurement of differential inclusive jet $p_{\rm T}$ spectra over
very wide $p_{\rm T}$ ranges, shown in Fig.~\ref{fig:jet}, is a very
strong test of model predictions. It has been shown that the shape of
the parton distribution functions is sensitive to such
distributions. In particular the gluon distribution could be improved
considering high-precision inclusive jet spectra. The very low-x gluon
parton distribution becomes more-and-more important when the
centre-of-mass energy increases, thus, it is relevant in
very high energy cosmic ray interactions. 

\begin{figure}
\includegraphics[width=\columnwidth]{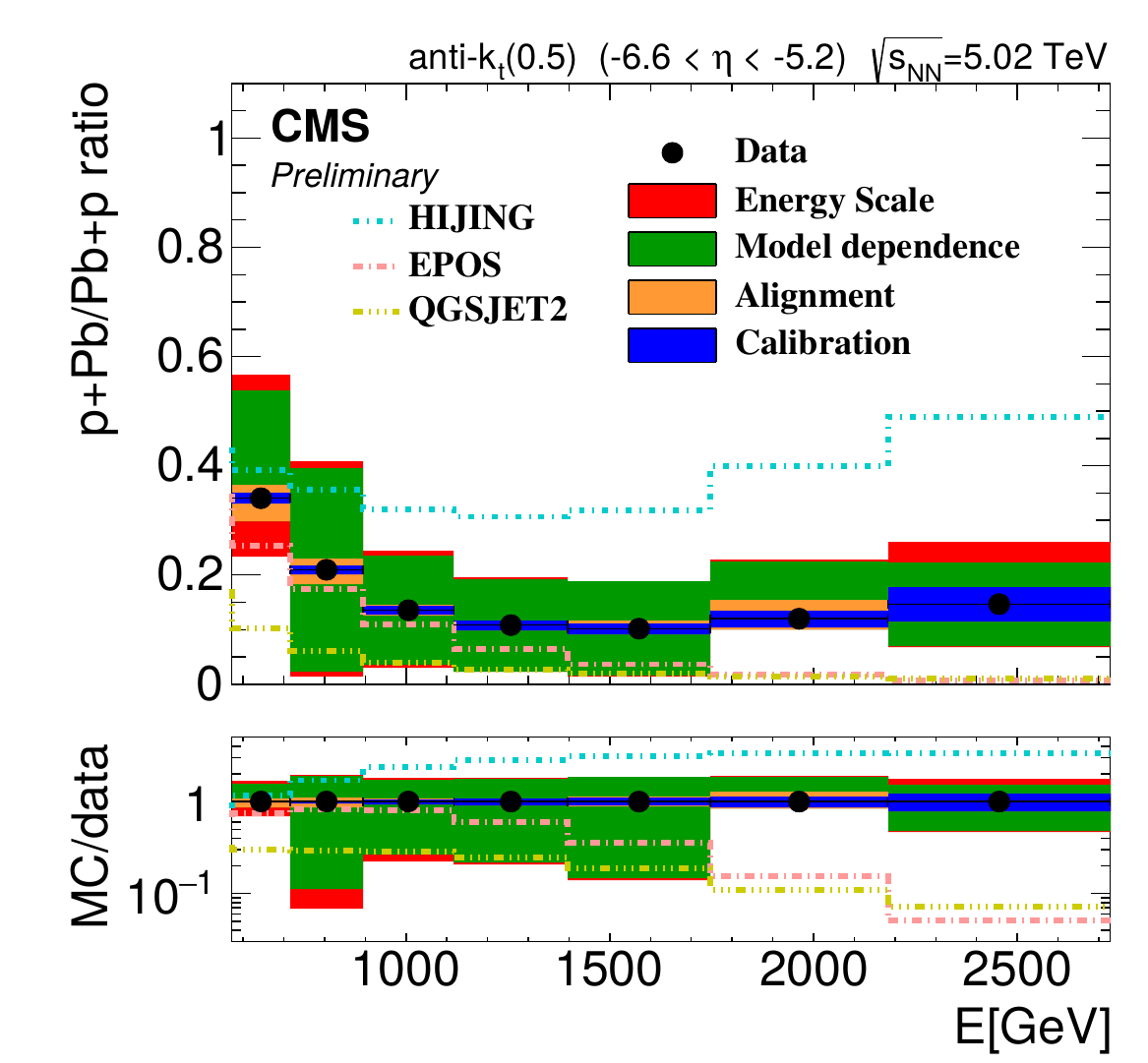}
  \caption{The ratio between proton-lead by lead-proton rates of very
    forward jets measured in CASTOR as function of the observed jet
    energy, see Ref.~\cite{Sirunyan:2018ffo}.}
\label{fig:pa}
\end{figure}

In cosmic ray air showers nuclear effects are another aspect that
requires precise modelling. The parton distribution functions in
nuclear interactions are subject to important further corrections,
like shadowing or saturation. This will also affect multi parton interactions and
the underlying event. For this reason it is very important that CMS
measured the very forward inclusive jet cross sections in proton-lead
collisions at $\sqrt{s_{\rm NN}}=5.02$\,TeV using data of the CASTOR
calorimeter, see Fig.~\ref{fig:pa}. These data highlights the large
importance of nuclear effects for very high energy cosmic ray
interactions.

\section{Underlying Event, multi parton interactions, double parton scattering}

\begin{figure}
\includegraphics[width=\columnwidth]{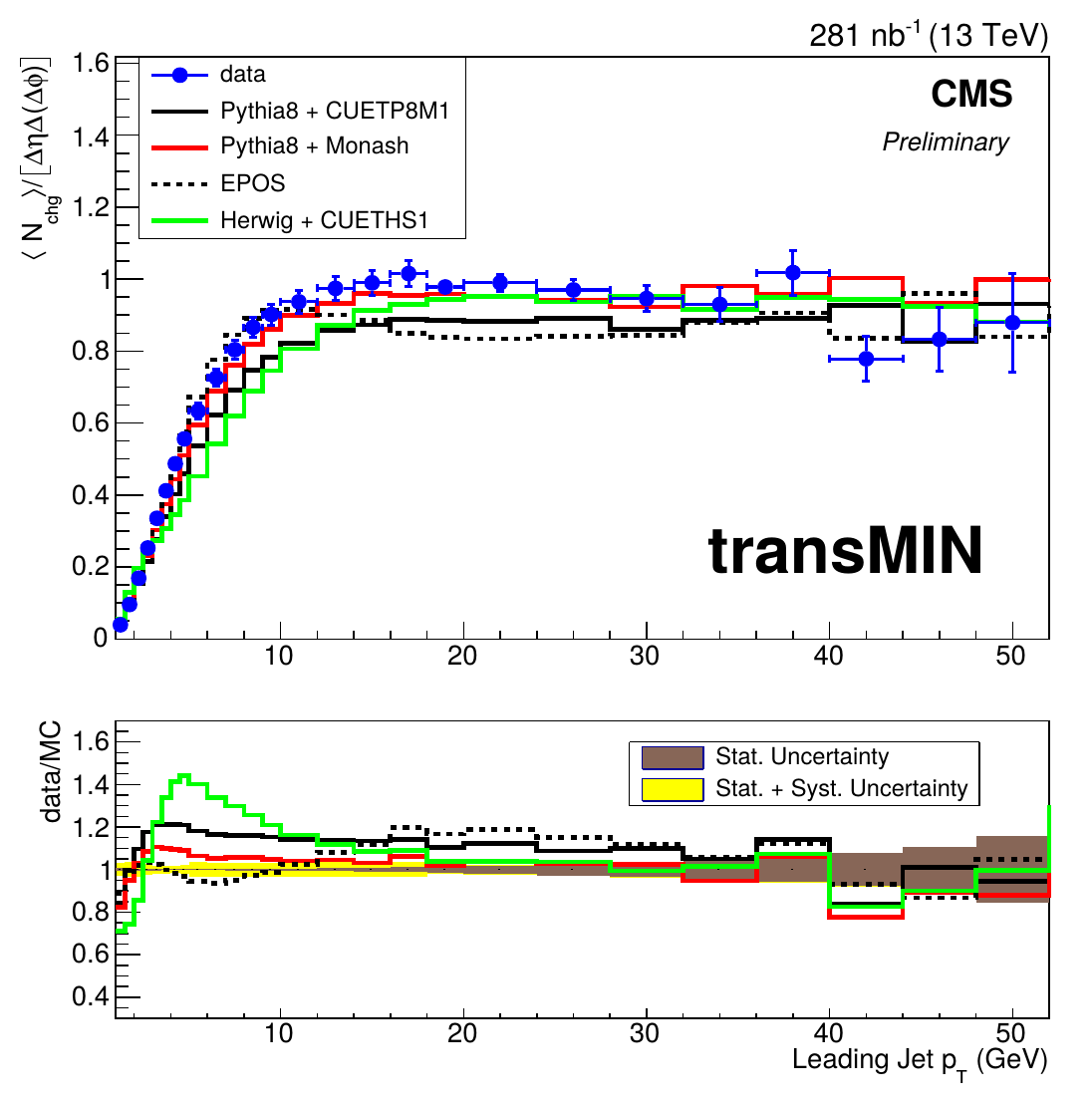}
\caption{The average number of charged particles as function of the
  leading jet $p_{\rm T}$ in proton-proton collisions at 13\,TeV, see
  Ref.~\cite{CMS:2015zev}. Shown here is the ``transMIN'' region,
  which is transverse to the leading jet and in the hemisphere with less activity. }
\label{fig:ue}
\end{figure}

The measurements of the underlying event is a typical experimental
method to assess the accuracy of QCD effects in event generation that
are very difficult to be determined from first principles. This includes
initial- and final-state radiation, multi parton interactions, and
projectile fragmentation. The corresponding measurement in
proton-proton collisions at 13\,TeV is shown in Fig.~\ref{fig:ue}. The
EPOS LHC model has some problems to describe the data in the region
around jet-$p_{\rm T}$ of 20\,GeV.

\begin{figure}
\includegraphics[width=\columnwidth]{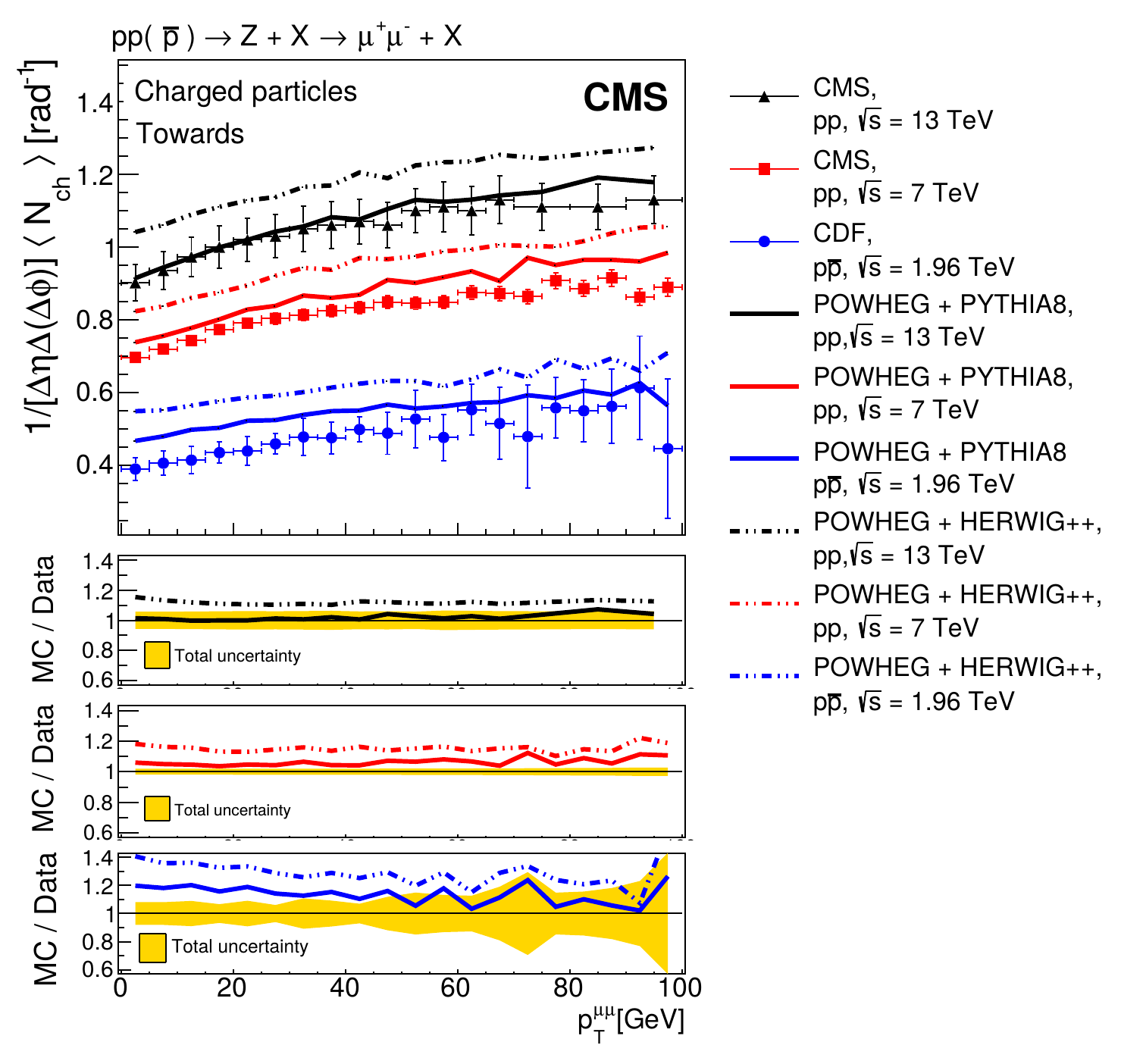}
\caption{The average number of charged particles as function of
  central Z $p_{\rm T}$, see Ref.~\cite{Sirunyan:2017vio}, in proton-proton collision of different centre-of-mass energies.}
\label{fig:ueZ}
\end{figure}

While typical underlying event measurements are performed relative to
the highest jet $p_{\rm T}$, CMS has also measured it in reference the
production of a Z boson~\cite{Sirunyan:2017vio}. This is shown in
Fig.~\ref{fig:ueZ}. 

\begin{figure}
\includegraphics[width=\columnwidth]{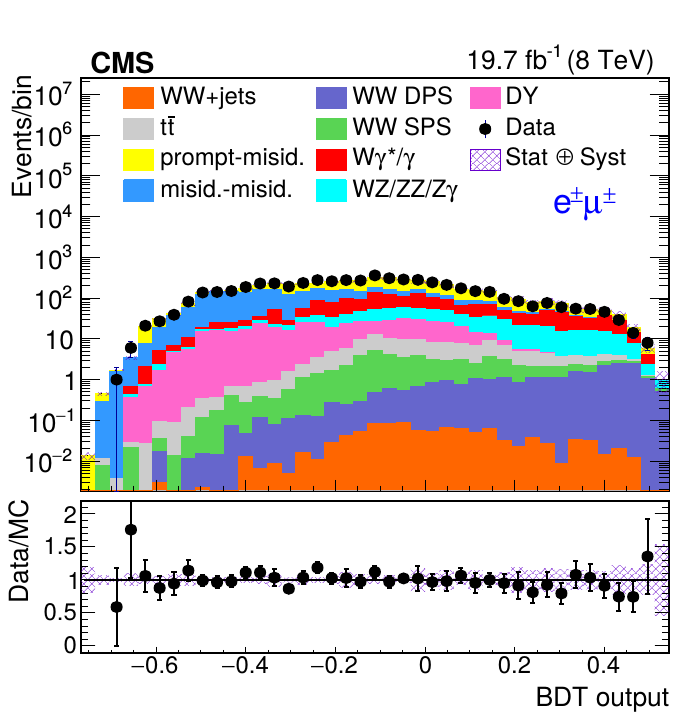}
\caption{Output of a boosted decision tree for same-sign WW events in
  proton-proton collisions at 8\,TeV. All relevant signal
  contributions are outlined. See Ref.~\cite{Sirunyan:2017hlu}.}
\label{fig:ssww}
\end{figure}

\begin{figure*}
  \centering
  \includegraphics[width=.9\textwidth]{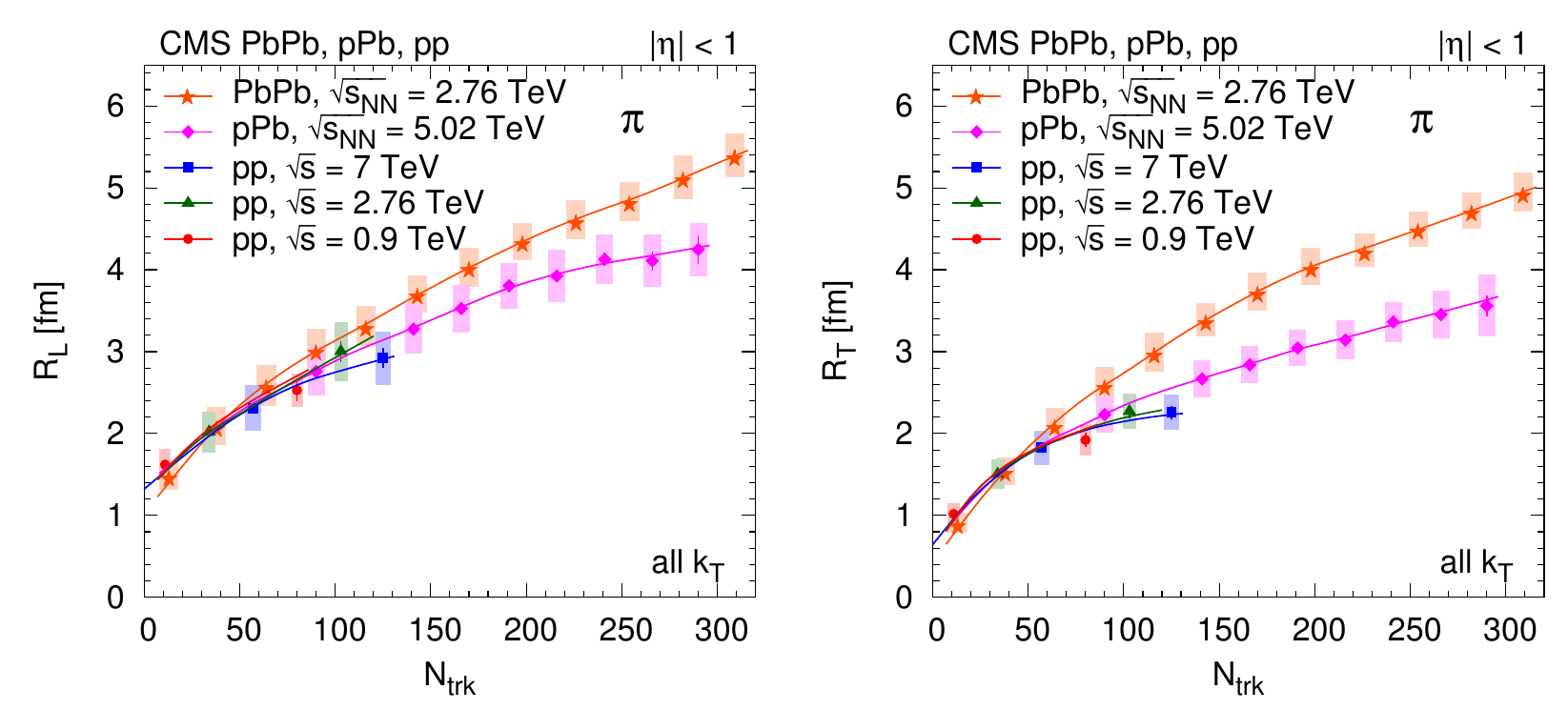}
  \caption{Geometry of the particle emission region measured in
    different collisions systems at different energies, see
    Ref.~\cite{Sirunyan:2017ies}. Shown are the parameters
    longitudinal size $R_L$ (left) and transverse size $R_T$ (right) as function of
    number of tracks, $N_{\rm trk}$.}
  \label{fig:bec}
\end{figure*}

Another way to study the internal partonic structure in hadron
collisions is to search for double parton scattering. In
Fig.~\ref{fig:ssww} the measurement of same-sign WW production in
proton-proton collisions at 8\,TeV~\cite{Sirunyan:2017hlu} is
shown. The backgrounds in such a measurement are numerous, thus, only a
multi-variate analysis is able to extract the quantity of interest.

\section{Correlations}

It is an important discovery at the LHC that secondary particle
production in proton-proton collisions contains a new type of
correlations that were not expected. This was also observed in
proton-proton collisions at 13\,TeV~\cite{Khachatryan:2015lva}. Such
two-particle correlations were only known from heavy-ion collisions
before. One possible conclusion from this could be that the underlying
mechanisms in proton-proton, proton-lead and lead-lead collisions are
the same and only depend on the maximum energy densities reached in
collisions.

Another option to look at correlations is to observe Bose-Einstein
correlations in the production of identical secondary particles. Due
to interference effects, the probability to observe identical
particles as a function of particle direction is modified. These data
will contain information on the geometry of the emission region. In
Fig.~\ref{fig:bec} some of these data are shown.

\section{Minimum bias results}

To describe very high energy cosmic ray interactions also minimum bias
data is of particular importance. In cosmic ray interactions there are
typically no filters or any sort of event selection. Thus the particle
production in typical minimum bias collisions is what is most
relevant. Here we will focus only on a single examples of this type of
measurements. 

\begin{figure}
\includegraphics[width=\columnwidth]{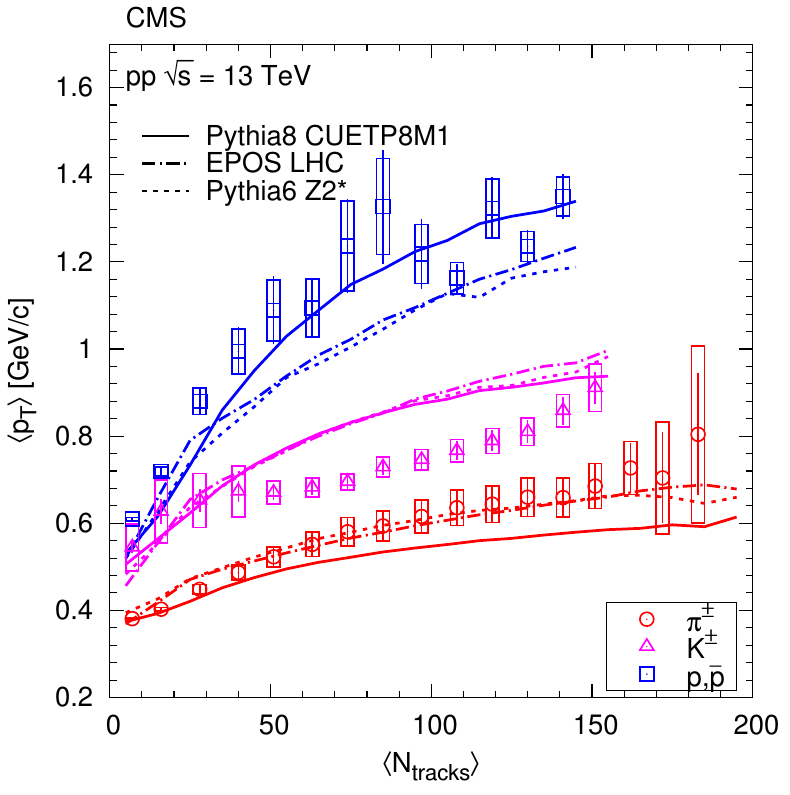}
  \caption{Measurement of the average $p_{\rm T}$ of identified
    particles versus the average number of charged tracks $N_{\rm
      tracks}$ in proton-proton collisions at 13\,TeV, see
    Ref.~\cite{Sirunyan:2017zmn}.}
\label{fig:pid}
\end{figure}

One of the directly relevant results for very high cosmic ray interactions is the detailed measurements of the
spectra of identified particles. In Fig.~\ref{fig:pid} some identified
particle spectra determined in proton-proton collisions at 13\,TeV are
shown. There is obvious room also by the EPOS-LHC model for
improvements.


\section{Summary}

A brief overview of some of the most recent CMS forward-physics
measurements has been reported. The physics covered by those
measurements covers a very wide and also diverse range of
fields. Several unique and very important observations relevant for
very high cosmic ray interactions are part of this.

\bibliography{rulrich.bib}
\end{document}